# Non-Euclidean interfaces decode the continuous landscape of graphene-induced surface reconstructions

Li-Qun Shen[1#], Hao-Jin Wang[1#], Mengzhao Sun[1#], Yang Xiang[2#], Xin-Ning Tian[1], Yue Chai[1], Yue Yang[1], Feng Ding[3], Xiao Kong[2], Marc-Georg Willinger[4], Zhu-Jun Wang[1*]

[1]School of Physical Science and Technology, ShanghaiTech University, Shanghai, China.
[2]Shanghai Institute of Microsystem and Information Technology, Chinese Academy of Sciences, Shanghai, China
[3]Suzhou Laboratory, Suzhou, Jiangsu, China
[4]Department of Chemistry, Technical University of Munich, Garching, Germany.

#These authors contributed equally to this work: Li-Qun Shen, Hao-Jin Wang, Mengzhao Sun, Yang Xiang
*E-mail: wangzhj3@shanghaitech.edu.cn

## Abstract

Interfacial reconstruction between two-dimensional (2D) materials and metal substrates fundamentally governs heterostructure properties, yet conventional flat substrates fail to capture the continuous crystallographic landscape. Here, we overcome this topological limitation using non-Euclidean interfaces—curved 2D graphene-copper surfaces as a model system—to traverse the infinite spectrum of lattice orientations. By integrating multimodal microscopy with a deep-learning-enhanced dimensional upscaling framework, we translate 2D scanning electron microscopy (SEM) contrast into quantitative three-dimensional (3D) morphologies with accurate facet identification. Coupling these observations with machine-learning-assisted density functional theory, we demonstrate that reconstruction is governed by a unified thermodynamic mechanism where high-index facets correspond to specific local minima in the surface energy landscape. This work resolves the long-standing complexity of graphene-copper faceting and establishes non-Euclidean surface topologies as a generalizable paradigm for decoding and controlling interfacial reconstruction in diverse metal–2D material systems.

## Introduction

The surface reconstruction of metals covered by 2D materials has been a focal point of surface science and materials engineering for the last decade[1-6]. In systems such as chemical vapor deposition (CVD) growth[7-10], the interaction between the 2D overlayer and the metal substrate strongly modulates interfacial atomic diffusion and arrangement[11-19]. Understanding the mechanisms behind these phenomena is critical for controlling growth quality, tuning surface reactivity, and designing low-dimensional heterostructures. However, despite extensive study, there remains no mature methodology to systematically derive reconstruction outcomes—particularly regarding high-index facets—for arbitrary combinations of metals and 2D materials. Previous works have largely relied on a case-by-case methodology using flat, single-crystal substrates[20-24]. Because a flat surface represents only a single crystallographic orientation, these studies capture only a fragmented view of the reconstruction landscape, leading to misaligned findings and even controversial interpretations of facet rearrangement[25-27].

The fundamental barrier to fully resolving these reconstruction mechanisms is the conflict between the discrete lattice indices exposed on a conventional polycrystal and the continuous nature of crystallographic orientation space. Topologically, the set of all possible facet indices is homotopic to a sphere in three-dimensional real space. Theoretically, as one increases the number of facets on a polyhedron, it converges toward a spherical limit; however, this limit is unreachable using standard Euclidean (flat) substrates. A planar sample represents only a tangent space at a single point on this theoretical sphere, making it mathematically impossible to access the general reconstruction landscape using ordinary metal foils or films.

To overcome this topological limitation, we introduce the use of non-Euclidean interfaces—specifically, curved surfaces with non-zero Gaussian curvature—to resolve the full orientation distribution of reconstructed facets. The curvature of the substrate surface has been applied to modulate the chirality of screw growth of 2D Van de Waals materials[28]. Unlike flat planes, these non-Euclidean surfaces possess continuous normal vector fields, allowing them to physically represent a dense, effectively infinite group of facet indices simultaneously. By utilizing a substrate that naturally mimics the spherical geometry of orientation space, we can observe the reconstruction behavior of the entire crystallographic spectrum in a single experiment.

Here, we utilize the graphene-copper (Gr-Cu) system as the paradigmatic "cutting point" to decode this universal surface reconstruction issue. We fabricated raised, graphene-covered Cu mounds that serve as non-Euclidean interfaces, displaying strong facet reconstruction across a continuous range of curvatures. To analyze this complex morphology, we developed a multimodal, multiscale framework integrating in situ scanning electron microscopy (SEM), electron backscatter diffraction (EBSD), and focused ion beam transmission electron microscopy (FIB-TEM). Furthermore, we employed a Poisson-based geometric modeling approach to upscale 2D SEM contrast data into quantitative 3D surface morphologies, validated by electron-optical simulations and scanning probe microscopy (SPM).

By correlating these experimental observations with high-throughput density functional theory (DFT) and machine-learning-based interatomic potentials, we demonstrate that the reconstruction is fundamentally energy-driven. We identified that specific high-index facets—such as {113}, {610}, {531}, and {711}—correspond to local minima in the potential energy surface. This work not only resolves the specific orientation selectivity of the Gr-Cu interface but also establishes a generalizable framework using non-Euclidean surfaces to explore the thermodynamic stability and reconstruction limits of diverse metal–2D material systems. Eventually, our methodology will promote researches about controlling interfacial interactions, tuning surface reactivity, and designing low-dimensional heterostructures.

## Orientation-Dependent Graphene-Induced Reconstruction on Non-Euclidean Cu Interfaces

Graphene growth induces strongly orientation-dependent reconstructions on Cu surfaces.

As illustrated in **Figs. 1a–f**, distinct crystallographic orientations manifest as specific step morphologies, scales, and alignments after graphene coverage. While previous studies utilizing conventional flat substrates have widely reported the exposure of low-index facets (e.g., {100}, {110}, and {111}), their Euclidean planar geometries capture only discrete, disconnected points in the space of all crystallographic orientations[20,23,24,29-31]. Consequently, although a vast number of high-index facets emerge in realistic systems, their exact indices and spatial distributions have remained unsystematically identified due to such topological limitation of flat surfaces.

To comprehend the complete landscape of reconstructed facets, we employed a stereographic projection sphere to map the continuous crystallographic orientation space. This sphere is the limit of infinitely cleaved polyhedron (**Fig. 1g**), reflecting the denseness and continuousness of lattice indices. We defined the spatial solid angle regions adjacent to three typical low-index reference planes—{100}, {110}, and {111}—as α, β, and γ, respectively (**Fig. 1h**). Considering the full m-3m point group symmetry of the face-centered cubic (FCC) copper lattice, applying rotational ($C_4$, $C_3$, $C_2$) and mirror symmetry operations to these fundamental regions generates a set of symmetry-equivalent domains (**Fig. 1g, right**). If these domains collectively cover the unit sphere with compatible gapless overlaps-which is in principle possible due to compactness of three-dimensional sphere $S^2$-the full spectrum of possible reconstructed facets can be systematically mapped (**Figs. 1i-j**).

Due to the topological constraints of Euclidean planes, flat surfaces are insufficient to represent such continuous region on the stereographic projection sphere. To overcome this barrier and physically accessing this continuous orientation spectrum, we fabricated non-Euclidean interfaces—specifically, curved surfaces with non-zero Gaussian curvature—by exploiting the suppression of Cu sublimation by graphene at high temperatures[13,32-35]. As shown in the *in situ* SEM observations in **Fig. 1k**, graphene nucleates on a pre-melting Cu (111) surface at 1000 °C and expands radially at a constant velocity $v$[36-38]. The graphene overlayer effectively shields the underlying copper from evaporation[13,34,39], meaning that for each radial position $r$, sublimation occurs only before the graphene front reaches it at $t = r / v$. Consequently, while the covered regions retain their height, the uncovered regions recede at a constant sublimation rate $E$[34]. This differential evolution creates a geometric profile $h(r, t)$ described by:

$$h(r, t) = h_0 - E \cdot min\left(t, \frac{r}{v}\right)$$

This process naturally yields a smooth, axisymmetric spherical-cap topology (**Figs. 1k$_1$-k$_3$**). Unlike flat samples restricted to a single tangent plane, this non-Euclidean interface possesses a continuous normal vector field, effectively allowing a single experimental sample to traverse a dense, continuous range of initial lattice orientations.

During the cooling process, this continuous non-Euclidean surface undergoes reconstruction[40]. As the spherical cap expands to cover a sufficiently large solid angle, the continuous curvature forces the surface to break symmetry and selectively expose thermodynamically stable facets along the sloped regions. This results in the formation of well-aligned, highly ordered step bunches composed of high-index facets, transforming the smooth spherical-cap dome into a polygonal faceted non-Euclidean mound (**Figs. 1k$_4$-k$_5$, l, m**). Ultimately, these distinct faceted structures, exhibiting clear crystallographic symmetries consistent with the underlying {100}, {110}, and {111} orientations, were experimentally captured, demonstrating that the non-Euclidean interface successfully decodes the continuous landscape of surface reconstruction in a single experiment.

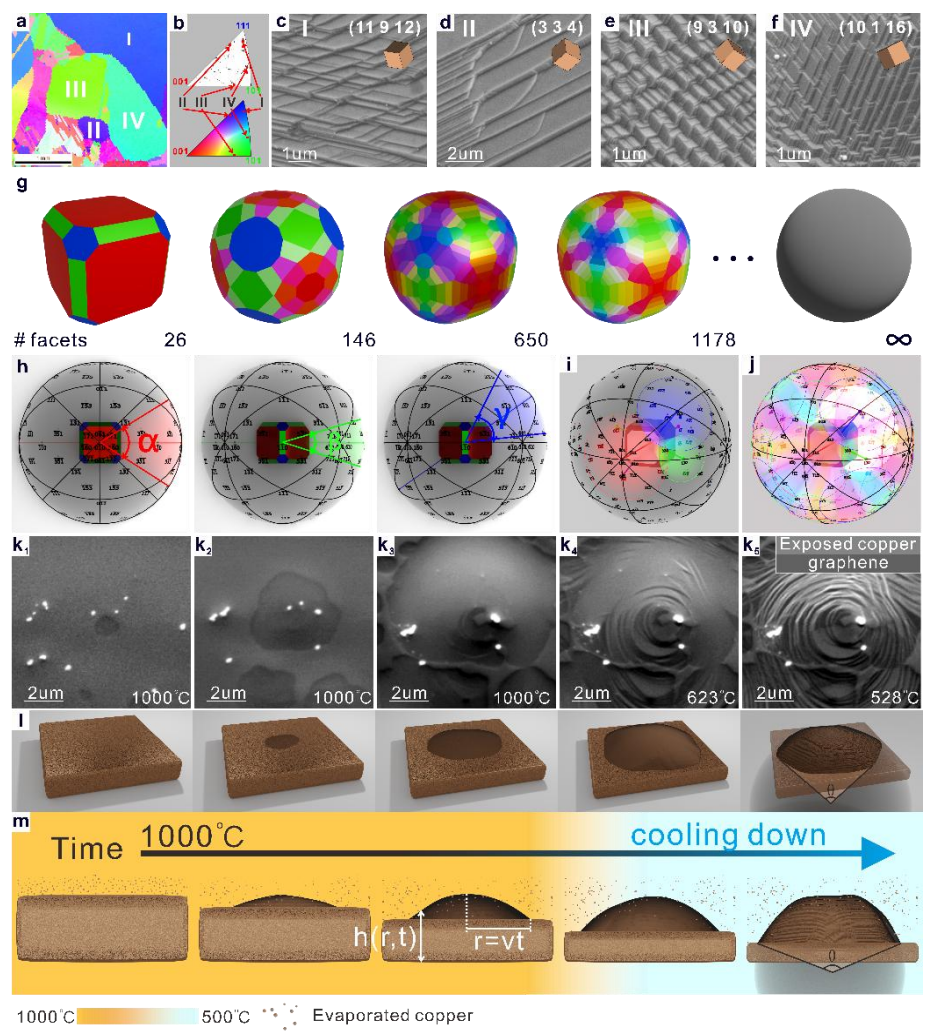

**Figure 1 | Orientation-dependent surface reconstruction on non-Euclidean Cu interfaces.**
**(a)** EBSD orientation map of a polycrystalline Cu surface showing four representative regions with distinct crystallographic orientations. **(b)** Corresponding EBSD inverse pole figure (IPF) key. **(c–f)** SEM images of the four regions marked in **(a)**, displaying orientation-dependent step morphologies and alignments on conventional flat (Euclidean) substrates. Insets: local crystallographic orientations. **(g)** Illustration of the topological relationship between discrete polyhedra and a continuous sphere. As the number of facets increases towards infinity (∞), the discrete polyhedron converges to a continuous spherical limit, highlighting that a flat surface represents only a tangent plane unable to access the full orientation space. **(h–j)** Theoretical framework for mapping reconstructed facets on a continuous sphere. **(h)** Definition of spatial solid-angle regions (α, β, γ) adjacent to low-index reference planes ({100}, {110}, {111}). **(i, j)** Generation of symmetry-equivalent regions by applying the full m-3m point-group symmetry, collectively covering the entire unit sphere to map all possible reconstructed facets. **($k_1$–$k_5$)** *In situ* SEM time-lapse observation of graphene growth and surface reconstruction on Cu(111). **($k_1$–$k_3$)** Nucleation and radial expansion of graphene at 1000 °C, where the suppression of Cu sublimation creates a smooth spherical-cap topology. **($k_4$, $k_5$)** Surface reconstruction during the cooling process, showing the transformation of the smooth dome into a faceted non-Euclidean mound composed of high-index planes. **(l)** Three-dimensional schematic evolution of the formation of the spherical-cap (non-Euclidean) interface. **(m)** Cross-sectional schematic and timeline of the surface evolution. The differential evolution creates a geometric profile $h(r,t)$ determined by the graphene expansion velocity ($v$) and Cu sublimation rate ($E$). The cooling step (Black to blue gradient arrow) illustrates the transformation from a smooth non-Euclidean dome into a reconstructed polygonal faceted non-Euclidean mound.

## Workflow for Facet Index Determination

In FCC systems, the {100}, {110}, and {111} projection planes exhibit fourfold ($C_4$), twofold ($C_2$), and threefold ($C_3$) rotational symmetries, respectively. SEM observations reveal that the graphene-induced faceted non-Euclidean mound structures strictly follow these inherent symmetries of the underlying crystallographic planes, resulting in highly ordered and uniform geometries, as shown in the left panel of **Fig. 2a**. To visualize how these symmetries dictate the spatial arrangement of reconstructed facets, we project the head of Alexander (from an ancient Greek coin) onto the stereographic sphere (**Fig. 2a, right**). This visualization intuitively demonstrates how the characteristic repetitive patterns of the spherical-cap (**Fig. 2a, left**) map directly to the fundamental symmetry operations of the substrate.

Based on this symmetry relationship, the following works focus on analyzing the step structure within the fundamental symmetry region, as indicated by the purple transparent area in **Fig. 2a**. The step orientations in other regions can be derived by applying corresponding symmetry operations to, and exhibit chiral relationships with those in the fundamental region[41,42]. For the same single-crystal domain, analysis of a single mound is sufficient since multiple faceted non-Euclidean mound structures exhibit consistent step alignment directions due to the well-defined crystallographic geometric relationships.

Within the same single-crystal domain, the steps can be classified as primary and secondary: the primary steps are parallel to the low-index substrate plane, while the secondary steps intersect the primary at specific angles. Both step families are constrained by a common direction, known as the crystallographic zone axis, which corresponds to the shared normal of the intersecting planes. Consequently, the orientation of the step bunches is not randomly distributed but strictly dictated by the zone axis direction. Only when these geometric constraints are satisfied can the steps reconstruct in an ordered fashion, ultimately forming step bunches with well-defined symmetry and orientation selectivity.

The workflow for determining the crystallographic indices of the secondary steps is illustrated in **Figs. 2b–f**. **Fig. 2b** shows the SEM image of the reconstructed region, and **Fig. 2c** presents the corresponding EBSD map, which enables precise identification of the substrate plane indices and determination of the primary step orientation. Furthermore, high-resolution three-dimensional surface topological data of the identical area are obtained by AFM, as shown in **Fig. 2d**, allowing for quantitative characterization of the faceted non-Euclidean mound morphology. The corresponding surface height profile is provided in **Fig. 2e**, from which the approximate angle between the primary and secondary steps can be determined. However, due to the absence of precise zone axis information, the AFM measurement direction is only approximately perpendicular to the steps, resulting in inevitable angular deviations that limit the accuracy of determining the true orientation of the secondary steps.

To achieve higher measurement precision, FIB milling is performed at the blue marked location in **Fig. 2d** to extract a cross-sectional lamella perpendicular to the step direction. Subsequently, the zone axis orientation of the cross-section is determined by TEM combined with electron diffraction. The precise angle between the primary and secondary steps is then measured from TEM images, as shown in **Fig. 2f**. The consistency between AFM and TEM measurements validates the reliability of the approach, and by integrating the zone axis information with the geometric relationships, the crystallographic indices and structural parameters of the secondary steps are accurately confirmed.

With the known indices of the primary and secondary steps and their orientation relative to the copper substrate, the surface normal distribution of all step facets within the reconstructed region can be quantitatively deduced. For example, the SEM image in **Fig. 2g** is converted into the corresponding surface normal map in **Fig. 2h**, where different colors represent different facet normals.

Based on the lattice index information acquired through foregoing procedure, we apply an enhanced Poisson surface reconstruction method to obtain a continuous and high-precision 3D surface simply from 2D SEM images[43,44]. The core idea is to treat the surface as a height function $z(x,y)$, and estimate its gradient using the components of the normalized surface normals. Specifically, assuming the normal vector at each point is given by $\boldsymbol{n}(x,y) = (n_x, n_y, n_z)$, the local slope of the surface in the x and y directions is approximated as:

$$p(x,y) = -\frac{n_x(x,y)}{n_z(x,y)}, q(x,y) = -\frac{n_y(x,y)}{n_z(x,y)}$$

These expressions provide the partial derivatives of the unknown surface depth $z(x,y)$. To recover the full surface from these gradients, we solve the following 2D Poisson equation:

$$\nabla^2 z(x,y) = \frac{\partial p(x,y)}{\partial x} + \frac{\partial q(x,y)}{\partial y}$$

Here, $\nabla^2$ is the Laplacian operator, which measures how the surface curves at each point. The right-hand side, representing the divergence of the gradient field, effectively tells the solver where the surface should bulge or sink to be consistent with the original normals.

To ensure a unique and stable solution, we apply Neumann boundary conditions (i.e., assume no change in height across the boundary), and fix the surface height at one reference point, such as the center. The resulting solution $z(x,y)$ is then normalized for display[45]. The reconstructed surface based on the normal information in **Fig. 2h** is shown in **Fig. 2i**.

To further verify the reconstruction accuracy, electron trajectory simulations are performed based on a global ray-tracing approach within the Monte Carlo theoretical framework. After modelling the electric field distribution based on the geometry of SEM device, the scattering, secondary electron emission, and trajectories of incident electrons interacting with the complex step structure are comprehensively reproduced, providing the simulated SEM image as shown in **Fig. 2j**. The simulation takes experimental parameters, including acceleration voltage, incident angles, and material-specific secondary electron yield into account, ensuring high fidelity with experimental conditions.

By comparing the simulated image in **Fig. 2**j with the experimental SEM image in **Fig. 2g** in terms of surface morphology, boundary characteristics, and symmetry features, the accuracy of step index determination and the 3D reconstruction process is effectively validated.

This integrated reconstruction and crystallographic indexing framework can be further applied to systematically explore all possible reconstructed facets and their spatial distributions on graphene-covered FCC copper surfaces, and even general two-dimensional material-metal interfaces, providing experimental insights into surface reconstruction mechanisms and the formation of high-index facets.

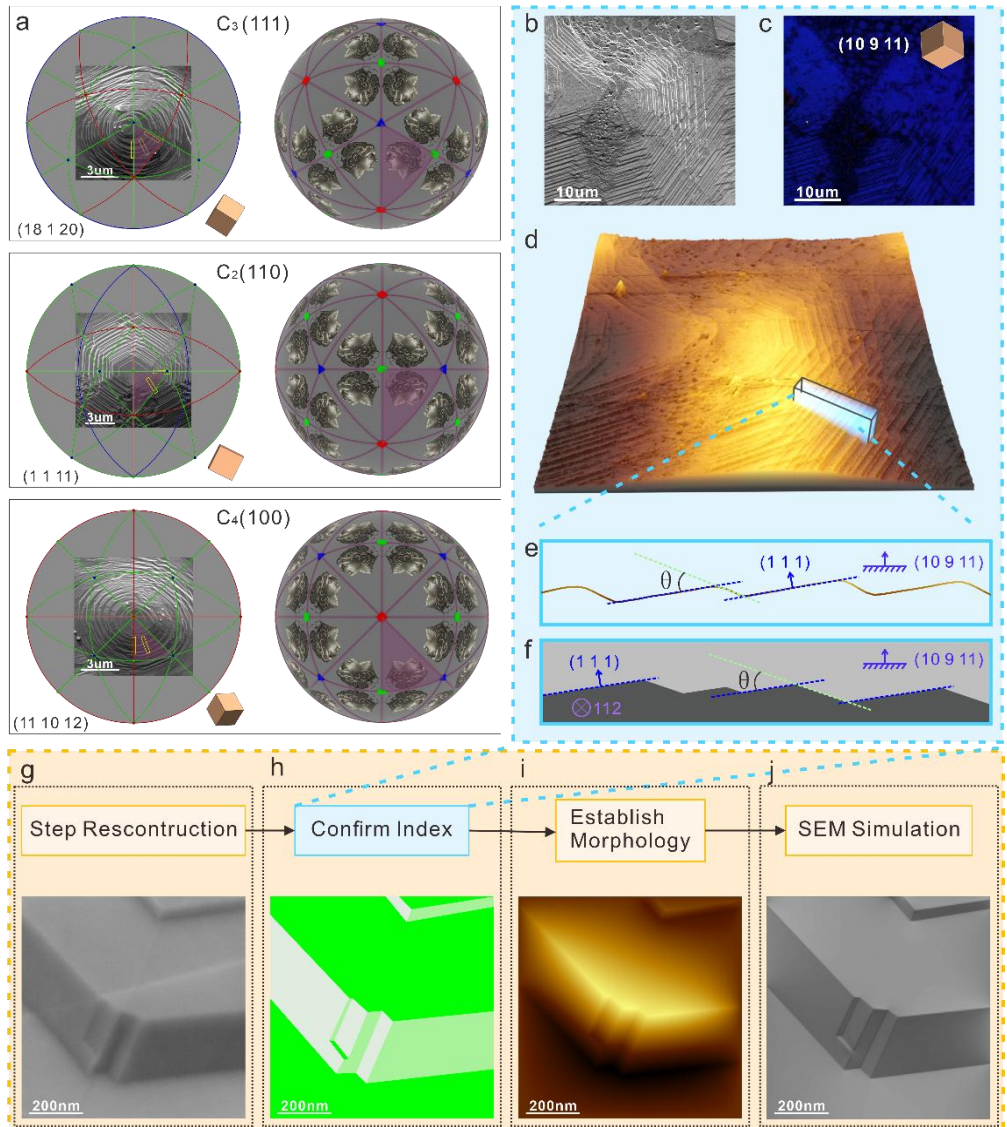

**Figure 2 | Multimodal framework for decoding crystallographic indices and 3D morphology of non-Euclidean interfaces.**

(a) Stereographic projections of Cu {111}, {110}, and {100} orientations illustrating their intrinsic $C_3$, $C_2$, and $C_4$ rotational symmetries. The projection of Alexander's head visually demonstrates how fundamental symmetry operations generate the equivalent orientations observed on the faceted non-Euclidean mounds (left: representative SEM images). **(b–f)** Experimental workflow for precise facet indexing. **(b)** SEM image of a target reconstruction region. **(c)** Corresponding EBSD map identifying the substrate orientation. **(d)** High-resolution AFM topography providing 3D surface constraints; the blue box indicates the FIB lift-out region. **(e)** AFM height profile showing the inter-step angle θ. **(f)** TEM cross-sectional analysis and simulated diffraction patterns used to rigorously determine the crystallographic zone axis and the exact inter-step angle, enabling unique facet identification. **(g–j)** Dimensional upscaling from 2D contrast to quantitative 3D morphology. **(g)** Experimental SEM image. **(h)** Derived surface-normal orientation map based on crystallographic data. **(i)** Quantitative 3D surface reconstructed from the normal vector field using an enhanced Poisson method, effectively recovering the non-Euclidean topology from 2D data. **(j)** Theoretical SEM image generated via Monte Carlo electron-trajectory simulation based on the reconstructed geometry. The high consistency with **(g)** validates the accuracy of the reconstruction framework.

## Decoding the Continuous Reconstruction Landscape via Symmetry and 3D Indexing

The transformation of smooth non-Euclidean domes into polygonal spherical-cap, as observed in the previous section, is not stochastic but rigorously governed by the intrinsic symmetry of the underlying crystal lattice. Reminding that the full stereographic projection sphere can be covered in a gapless way by three types of domains (α, β, γ) centered by the three low-indexed orientations, , we simply need to apply our step index determination workflow to the fundamental symmetric region of the non-Euclidean interfaces formed on Cu {111}, {110}, and {100} surfaces to thoroughly decode the complex reconstruction geometry of Graphene-Cu system. Taking the nearly dodecagonal faceted mound reconstructed on the Cu {111} surface as a paradigm, we show in details how the full orientation distribution can be resolved by identifying just a few key facets and exploiting crystallographic symmetry.

Unlike a flat sample that shows only one facet type, the curved non-Euclidean mound naturally breaks translation symmetry to expose multiple step families. The overall reconstruction landscape can be fully restored by identifying the secondary facet indices from three adjacent step bundles and applying the intrinsic $C_3$ symmetry of the {111} plane. Specifically, as shown in **Fig. 3a**, we selected three representative step regions (marked ①, ②, ③) within the fundamental symmetry domain of the faceted mound. To determine their exact crystallographic nature, we prepared cross-sectional lamellae along the step directions using focused ion beam (FIB) milling, with zone axes indicated by arrows.

The crystallographic orientation was rigorously determined through a two-stage diffraction analysis (**Fig. 3b**). First, Kikuchi patterns (**Fig. 3b$_1$**) provided a preliminary calibration of the zone-axis orientation, capturing the overall symmetry. However, to eliminate angular deviations caused by imaging conditions or local strain, we switched to selected-area electron diffraction (SAED) under identical tilt conditions. The resulting high-resolution diffraction patterns (**Fig. 3b$_2$**) enabled precise zone-axis calibration. By combining this exact orientation with the inter-facet angle θ measured directly from TEM images between the primary {111} terrace and the secondary step (**Fig. 3 b$_1$**), we uniquely identified the Miller indices of the secondary facets. Throughout the whole process, the angle between terrace and step in real space images is aligned with the AFM results, ensuring the zone axis obtained through tilting sample in the TEM is exactly the expected one before preparing cross-section samples (**Fig. 3b$_3$**).

The analysis revealed a distinct set of high-index facets: Region ① corresponds to a 〈110〉 zone axis with a {110} secondary facet; Region ② corresponds to a 〈112〉 zone axis with a {135} secondary facet; and Region ③ corresponds to a 〈110〉 zone axis with a {113} secondary facet.

Thus, the reconstruction on graphene-covered Cu {111} is dominated by the exposure of {110}, {135}, and {113} facets. For intuitive visualization, all measurement regions in **Figs. 3a–b** are color-coded according to the EBSD orientation key.

To validate this crystallographic model against the physical morphology, we overlaid the theoretically derived facet distribution onto the experimental SEM image (**Fig. 3c**). Note that chiral counterparts (e.g., {135} and {153}) are displayed in the same color for clarity, though their normal vectors remain distinct in our model. Leveraging the continuous normal vector field of the non-Euclidean surface, we performed Poisson surface reconstruction to recover the quantitative 3D morphology of the faceted mound (**Fig. 3d**). The accuracy of this reconstruction was verified by comparing it with AFM topography of the similar region (**Fig. 3f**). The cross-sectional profiles (**Fig. 3e**) show that the inter-facet angles derived from our Poisson model match the AFM measurements with a deviation of less than $\pm 0.8°$, confirming the high fidelity of our facet indexing and 3D reconstruction methodology.

We extended this workflow to other low-index orientations to demonstrate universality. For Cu {110} surfaces (**Figs. 3g–i**), the reconstruction exposes {111} and {135} secondary facets. For Cu {100} surfaces (**Figs. 3j–l**), it reveals {610} and {711} facets (). These findings provide a systematic and complete picture of graphene-induced high-index step reconstructions, proving that non-Euclidean interfaces can successfully decode the continuous orientation landscape that remains inaccessible to conventional flat substrates.

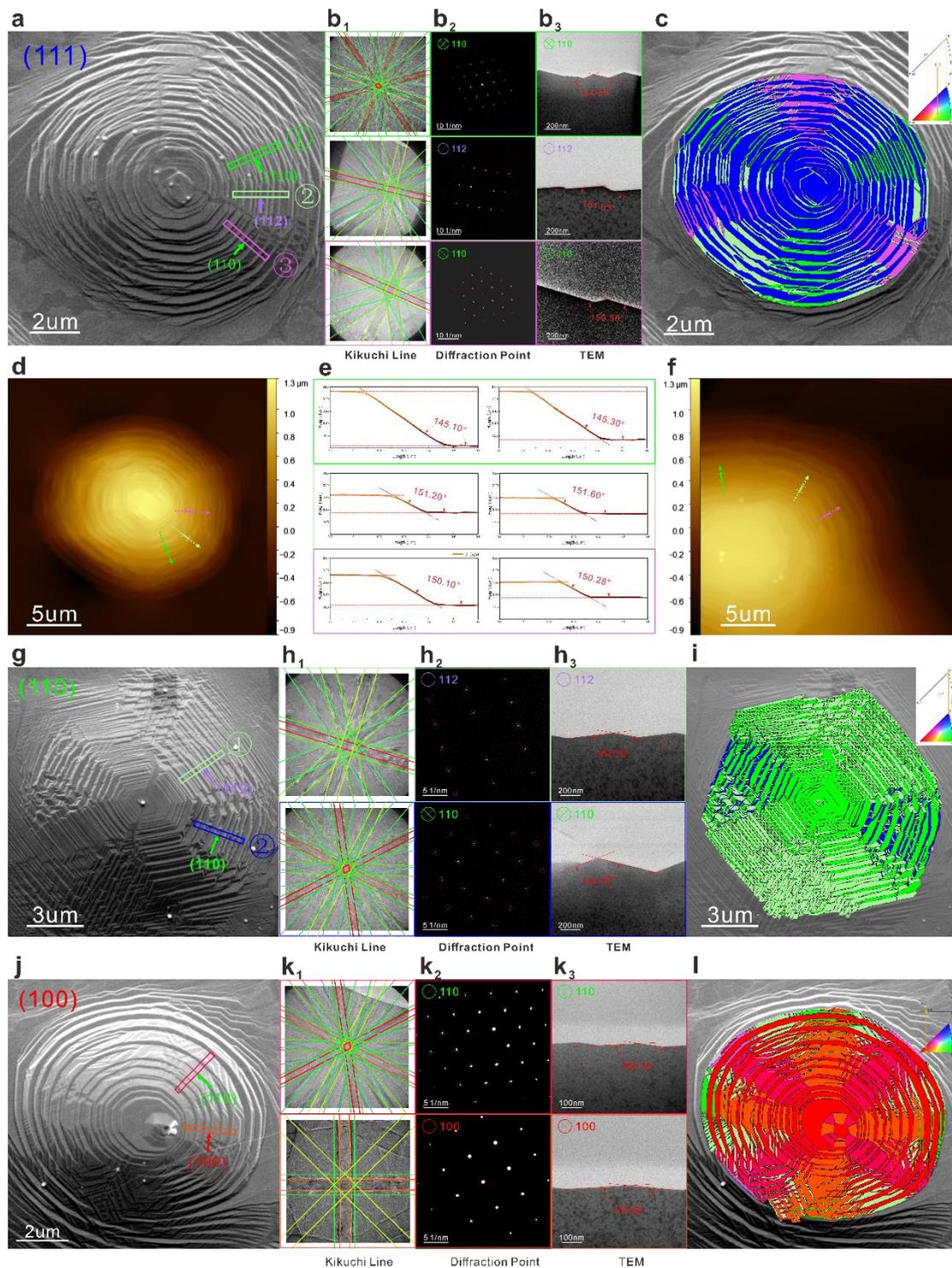

**Figure 3 | Decoding 3D morphology and facet indexing of graphene-induced faceted non-Euclidean mounds on Cu surfaces.**

(**a**) SEM image of a nearly dodecagonal faceted non-Euclidean mound reconstructed from the Cu{111} surface, showing three representative step regions (①–③). Arrows indicate the crystallographic zone axes of FIB-prepared cross sections. (**b₁–b₃**) TEM-based analysis workflow: (**b₁**) Kikuchi pattern for preliminary zone-axis determination; (**b₂**) high-resolution electron diffraction for precise orientation calibration; (**b₃**) TEM image for measuring the inter-facet angle between primary and secondary steps. The identified secondary facets correspond to {110}, {135}, and {113}. (**c**) Experimental SEM image overlaid with the theoretical facet distribution; different colors represent crystallographic orientations according to the EBSD color key. (**d**) 3D morphology reconstructed from surface-normal vectors using the Poisson reconstruction method. (**e**) Comparison of inter-facet angles obtained from AFM measurements and Poisson reconstruction, showing deviations below 0.5°. (**f**) AFM topography of the same region. (**g–i**) Analysis of the Cu {110} surface following the same workflow: SEM observation (**g**), TEM analysis (**h**), and comparison between experimental and theoretical facet maps (**i**), revealing secondary facets {111} and {135}. (**j–l**) Corresponding results for the Cu {100} surface, showing secondary facets {610} and {711}.Dashed arrows in (**d, f, j, l**) mark AFM scan directions, and solid lines correspond to the height profiles shown in (**e, k**), where color codes match the step regions indicated in the SEM images.

## Electron-Optical Validation via High-Fidelity Simulation

After obtaining the 3D morphology of the faceted non-Euclidean mounds, we validated the morphology by generating theoretical SEM images through electron-optical simulations that precisely replicate the experimental environment. As illustrated in **Fig. 4a**, an electron-optical model was constructed to achieve a geometric modelling of SEM instrument, accurately capturing the spatial configuration of the pole piece, secondary electron detector, and sample stage. The electrostatic field distribution (**Fig. 4b**) was then calculated based on these geometries, with beam parameters (acceleration voltage and current) set to match the experimental conditions exactly. Within this realistic field, we simulated the trajectories of secondary electrons emitted from the reconstructed surface to determine their collection efficiency by the detector.

This process yields the effective solid angle associated with each surface position (**Fig. 4c**)[46,47]. The total angular space is represented in spherical coordinates, where the reference axis is defined by the intersection between the detector–pole-piece plane and the horizontal plane. The white region indicates the range of electron emission that can be detected (**Fig. 4c**). The spatial distribution of these effective solid angles is shown in **Fig. 4d**. After accounting for surface shadowing effects, the detectable electron count at each point was calculated, directly reflecting the local image contrast. This allows reconstruction of simulated secondary-electron images corresponding to the 3D surface morphology. The strong agreement between simulated and experimental SEM contrast distributions provides direct validation of the reconstruction accuracy.

Based on this framework, we analyzed representative faceted non-Euclidean mound structures on Cu {100}, {110}, and {111} surfaces (**Fig. 4e**). Their 3D morphologies were reconstructed using the Poisson method (**Fig. 4f**), and the corresponding theoretical SEM images were obtained through electron-optical simulation (**Fig. 4g**). To quantitatively evaluate contrast consistency, grayscale intensity profiles were extracted from the same positions in the experimental and simulated images (indicated by white arrows) and compared (**Fig. 4h**). The two profiles exhibit nearly identical trends and feature correlations, with only minor smoothing in the experimental data due to instrumental noise and electron-scattering effects—consistent with expected measurement limits. These results confirm that the reconstructed and simulated morphologies of Cu {100}, {110}, and {111} faceted non-Euclidean mounds are highly reliable, demonstrating the robustness and general applicability of the proposed reconstruction–verification framework for complex crystallographic surfaces.

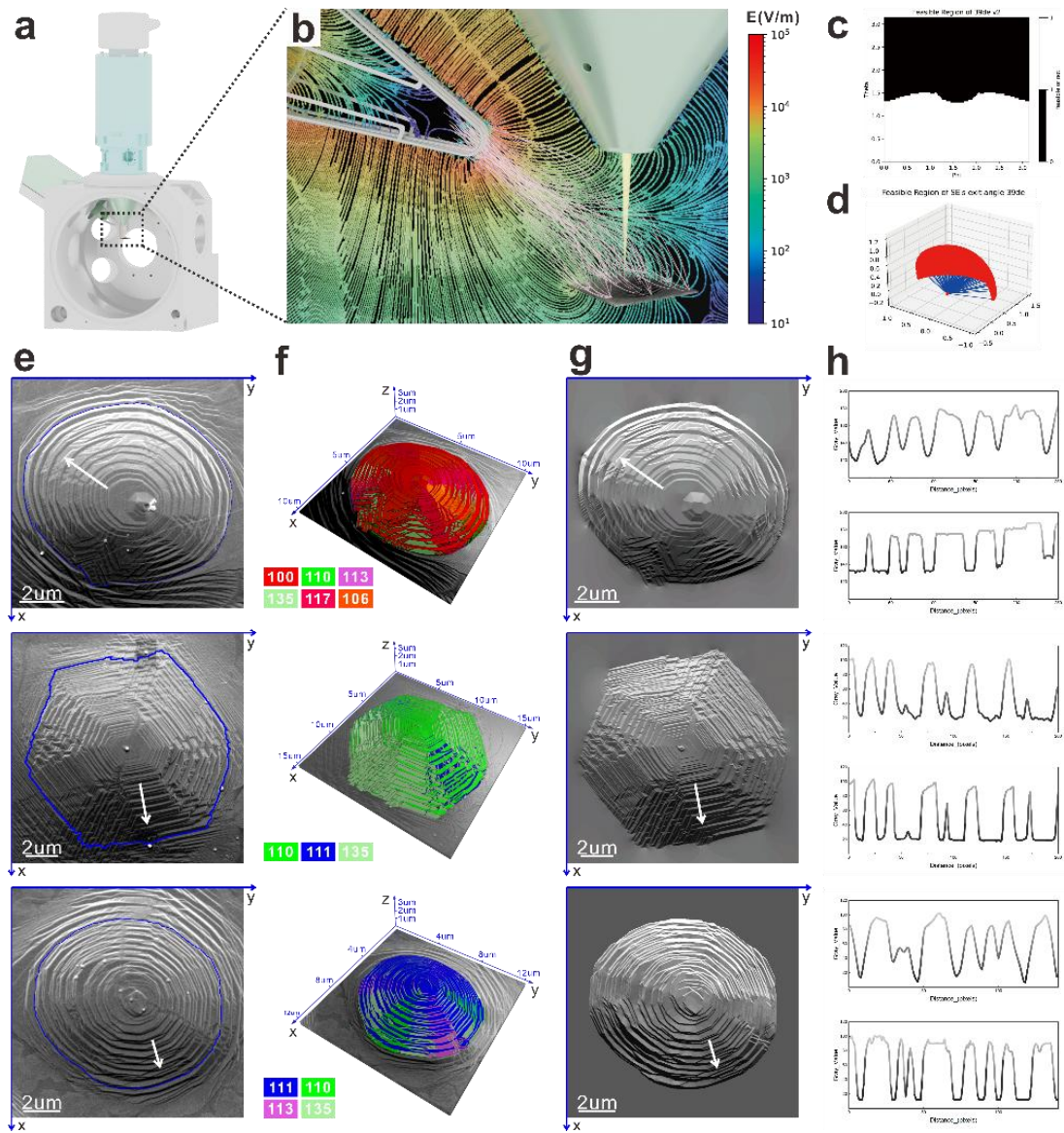

**Figure 4 | Rigorous validation of reconstructed non-Euclidean morphologies via high-fidelity electron-optical simulation.**

(a) Schematic of the electron-optical model constructed as a **1:1 geometric replica** of the experimental setup, incorporating the precise configuration of the pole piece, Everhart Thornley detector (ETD), and sample stage. (b) Finite-element simulation of the electric field distribution under boundary conditions and beam parameters (acceleration voltage and current) identical to the experiment. (c) Calculated effective solid angle for secondary-electron detection, visualized in spherical coordinates (white region indicates detectable emission angles). (d) Spatial distribution of detectable electron emission after considering surface-shadowing effects. (e–h) Validation across different crystallographic orientations. (e) Experimental SEM images of faceted non-Euclidean mounds on Cu {100}, {110}, and {111} surfaces. (f) Quantitative 3D morphologies of the corresponding regions recovered via Poisson surface reconstruction. Colors represent facet orientations (see inset legends). (g) Theoretical SEM images generated by tracing secondary electron trajectories within the realistic field environment shown in (b), based on the geometries in (f). (h) Comparative grayscale intensity profiles extracted along the white arrows in **(e)** (experimental) and **(g)** (simulated). The profiles exhibit nearly identical contrast modulation and periodicity, confirming the high reliability of the reconstruction–simulation framework.

**Uncovering the Thermodynamic Origin of Orientation Selectivity**

After experimentally and electron-optically verifying the orientations and geometries of the reconstructed facets, we further employed theoretical calculations to assess their thermodynamic stability and reveal the energy-driven mechanism underlying graphene-induced surface reconstruction. To validate the experimentally determined facet orientations, we performed systematic surface-energy calculations of graphene-covered Cu surfaces using density functional theory (DFT) combined with machine-learning interatomic potentials.

Based on our calculation results, the reconstruction of graphene-covered metal surfaces is primarily driven by the minimization of total system energy. While mechanical stress is a potential factor, our analysis indicates that the dominant mechanism is step bunching[17], rather than stress relief (see Supplementary Information). This process involves a trade-off between two energetic contributions: (1) the energy penalty from graphene bending across steps, and (2) the energy gain from transforming high-energy surfaces into lower-energy configurations. At experimentally accessible scales, we found the bending contribution to be negligible. Consequently, the reconstruction can be quantitatively described by a Wulff construction[48,49], where the system achieves a global energy minimum when the exposed facets coincide with thermodynamically stable planes on the Wulff plot.

Accurately modeling this interface presents a significant computational challenge. The intrinsic lattice mismatch between graphene and Cu varies with crystal orientation—reaching ~3.5% for Cu (111)[50]. Conventional DFT methods often impose artificial compression on the substrate to enforce "perfect epitaxy," which neglects the large-period moiré patterns observed experimentally and introduces spurious strain energy. To overcome this, we developed high-precision machine-learning potentials trained on extensive datasets. These models accurately predict both the surface energy of anisotropic Cu orientations (root mean square error (RMSE): 3.22 meV/atom) and the Cu–graphene interfacial binding (RMSE: 3.20 meV/atom), offering near-DFT accuracy at a fraction of the computational cost (**Fig. 5a**).

Utilizing these efficient machine-learning potentials, we developed a generation algorithm for near-commensurate van der Waals heterostructures, keeping substrate strain below 2%. This enabled high-throughput energy calculations across hundreds of orientation systems. The reliability of this framework was confirmed by the low deviation (<5%) from full DFT benchmarks (**Fig. 5b**). We then mapped these theoretical surface energies onto the stereographic projections of Cu (111), (110), and (100) facets, directly mirroring the experimental non-Euclidean sectors. To ensure coverage of all plausible step orientations, we applied a ±3° angular tolerance (**Fig. 5d**). The resulting three-dimensional energy landscapes (**Fig. 5e**) reveal a striking correlation: the local energy minima (marked in red) precisely match the experimentally identified facets — {111}, {110}, {100}, {113}, {610}, {531}, and {711}.

This perfect alignment confirms that the specific high-index facets emerging on the non-Euclidean mounds are indeed thermodynamically stabilized by the graphene overlayer. Furthermore, by reconstructing the theoretical Wulff morphology based solely on calculated energies (**Fig. 5f**), we reproduced the experimentally observed geometries with high fidelity.

Crucially, the experimental facet determination and theoretical energy mapping were conducted independently, without parameter fitting. This consistency constitutes a rigorous double-blind test between experiments and theory, demonstrating that graphene-induced Cu surface reconstruction is fundamentally an energy-driven, orientation-selective process governed by universal thermodynamic principles.

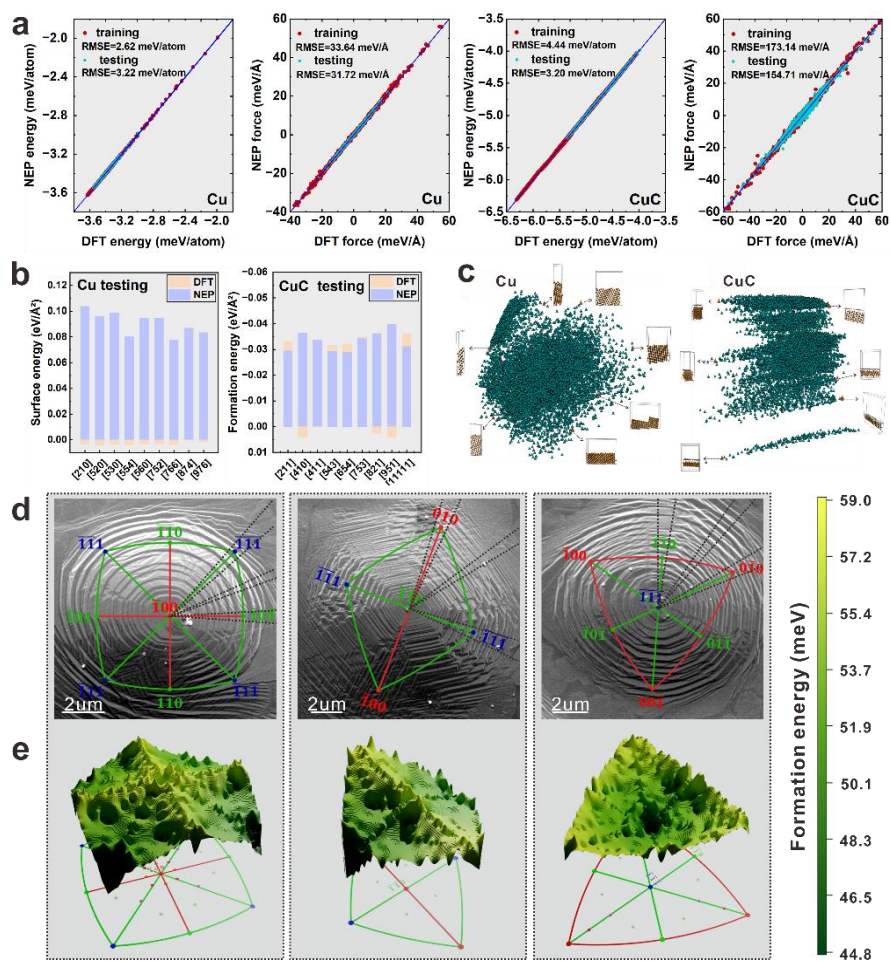

**Figure 5 | Theoretical validation of graphene-induced Cu surface reconstruction via machine-learning–accelerated surface-energy mapping.**

(a–c) Development and validation of high-precision machine-learning interatomic potentials. (a) Parity plots comparing machine-learning predictions with density functional theory (DFT) ground truth, demonstrating excellent accuracy for both Cu and Cu–graphene (CuC) systems (Energy RMSE: ~3.2 meV/atom; Force RMSE: <155 meV/Å). (b) Comparison of DFT and ML-predicted surface and interface energies for selected low-index Cu facets and Cu–graphene interfaces, confirming high predictive accuracy (< 5% deviation). (c) Structural sampling distributions for Cu and CuC configurations used in model training, ensuring broad coverage of high-index surfaces and interfacial geometries. (d) Experimental SEM images of reconstructed Cu surfaces with {100}, {110}, and {111} orientations, annotated with experimentally determined step directions and facet indices. (e) Computed 3D surface-energy landscapes corresponding to the stereographic regions in (d). Energy magnitude is represented by both color and height. The lower panels show the associated stereographic projections, where red dots mark local energy minima that correspond to experimentally identified stable high-index facets ({111}, {110}, {100}, {113}, {610}, {531}, and {711}). **(f)** Wulff constructions derived from the computed energy landscapes, showing thermodynamically stable facets and their spatial arrangements. The theoretical morphologies reproduce the experimentally observed faceted geometries, confirming that graphene-induced Cu surface reconstruction is an energy-driven, orientation-selective process.

### Generalization to Arbitrary Orientations and Global Completeness

To demonstrate the universal applicability of our spherical-cap-based framework beyond specific non-Euclidean mounds, we applied the established reconstruction models (for {111}, {110}, and {100} sectors) to analyze arbitrary Cu surfaces with random orientations. We revisited the representative regions shown in **Figs. 1c–f** (detailed in **Figs. 6a–d**). First, the substrate orientations were defined by EBSD. These arbitrary orientations were then mapped onto our established low-index spherical-cap models via stereographic projection, enabling the predictive identification of secondary facets based on their geometric location.

For instance, **Fig. 6e** and **6f** correspond to the {111} spherical-cap model. Here, the facet orientations of the reconstructed regions in **Fig. 6a** and **6b** are marked by an orange triangle and a red circle, respectively. Notably, the unstable orientation marked by the orange triangle energetically relaxes toward the three nearest local minima. This decomposition into specific stable high-index facets is accurately traced by the yellow arrows, explaining the multi-faceted morphology observed in real space. Similarly, the regions in **Fig. 6c** and **6d** were successfully mapped onto the {110} spherical-cap model (**Figs. 6g, h**).

Guided by this theoretical scheme, we assigned the indices of these three predicted energy minima to the corresponding facets observed in the SEM images, generating the spatially resolved orientation maps shown in **Figs. 6i–l**. To quantitatively validate these predictions, we performed AFM 3D topography measurements at the exact same locations. The inter-step angles derived from AFM show excellent agreement with our theoretical predictions (see Supplementary Information for profiling details). This consistency confirms that our framework, though derived from non-Euclidean mounds, can accurately resolve and predict reconstructed facets on flat Cu surfaces of any arbitrary orientation.

Finally, we assessed the topological completeness of our model. We projected the three representative low-index faceted non-Euclidean mounds onto a unit sphere to define their occupied solid-angle domains (**Fig. 6n**). Applying the full $O_h$ (m-3m) point-group symmetry operations of the FCC copper lattice, we generated all symmetry-equivalent orientations. As visualized in **Fig. 6o**, these domains collectively form a gapless tessellation that fully covers the unit sphere. This proves that our reconstruction model is mathematically comprehensive for graphene-induced reconstructions on FCC Cu under low-oxygen conditions, excluding scenarios where oxygen drives distinct surface reconstructions[51-56]. Essentially, any stepped facet observed on a graphene-covered Cu surface falls within one of these symmetry-equivalent units, allowing its Miller index to be uniquely determined.

Finally, to schematically visualize the overall distribution of reconstructed facets, the model was unfolded onto a Wulff construction referenced to the {111} direction **(Fig. 6p)**. This representation clearly illustrates how graphene-induced reconstruction promotes the selective exposure of high-index facet families adjacent to {111}, {110}, and {100} orientations, providing a symmetry-complete map of surface reconstruction on Cu.

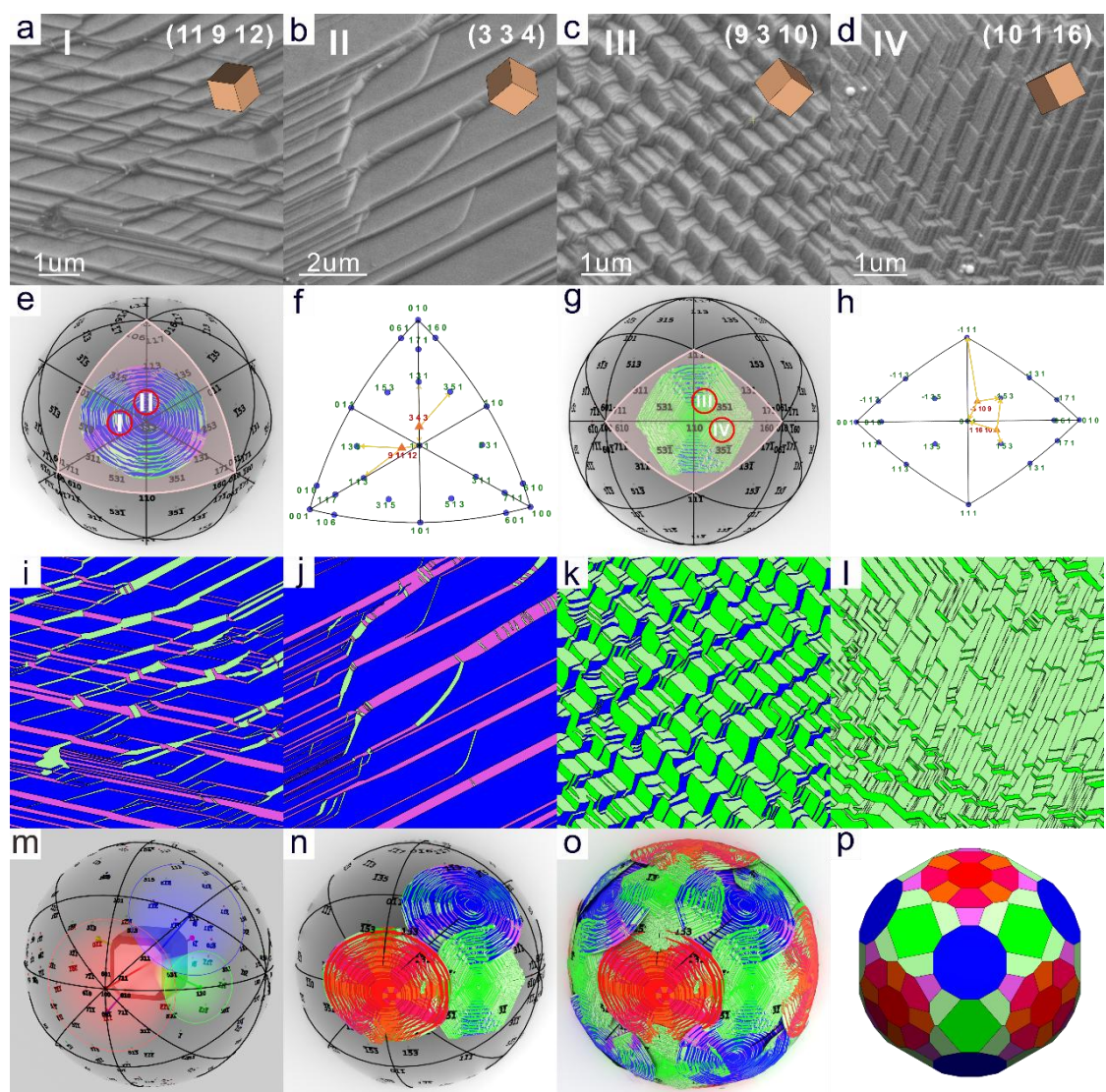

**Figure 6 | Verification and completeness of the spherical-cap-based facet identification framework.**

**(a–d)** SEM images of arbitrary Cu surfaces with random orientations (revisiting regions I–IV from **Figs. 1c–f**), used to test the model's generalization capability. **(e–h)** Predictive mapping of these arbitrary orientations onto the established spherical-cap models. **(e, f)** Mapping of regions I (**a**) and II (**b**) onto the {111} sector. The orange triangle (**f**) marks an unstable orientation that energetically relaxes toward three local minima (yellow arrows), explaining the multi-faceted decomposition observed in real space. **(g, h)** Mapping of regions III (**c**) and IV (**d**) onto the {110} sector. **(i–l)** Spatially resolved orientation maps generated by assigning the theoretically predicted facet indices to the experimental SEM images. The accuracy of these assignments was quantitatively validated by AFM profiling (see Supplementary

Information). **(m–p)** Demonstration of topological completeness. **(m, n)** Projection of the three representative faceted non-Euclidean mounds onto a unit sphere to define their fundamental solid-angle domains. **(o)** Generation of a **gapless tessellation** covering the entire unit sphere by applying full $O_h$ (m-3m) symmetry operations, proving that the model captures all possible graphene-induced reconstructions on FCC Cu. **(p)** A **symmetry-complete atlas** of reconstructed facets visualized on a Wulff construction (referenced to {111}), highlighting the selective exposure of stable high-index families across the global orientation space.

## Conclusion

In summary, this work overcomes the topological limitations of conventional flat substrates by exploiting non-Euclidean interfaces to decode the overall continuous landscape of graphene-induced surface reconstructions. By converting the flat metal surface to a curved space with non-zero Gaussian curvature, we transformed the problem of infinite crystallographic orientations into a tractable, continuous geometric system. This allowed us to systematically traverse the entire orientation spectrum within several samples, resolving the complete set of high-index facets that emerge on FCC copper surfaces.

Experimentally, we established a multimodal framework integrating EBSD, AFM, and FIB-TEM to determine precise crystallographic constraints. We further developed a dimensional upscaling method using Poisson surface reconstruction, which recovers quantitative 3D morphologies directly from 2D SEM contrast. The accuracy of this digital-analogy approach was rigorously validated by high-fidelity electron-optical simulations, demonstrating a reliable pathway for extracting 3D crystallographic information from general microscopy images.

Theoretically, we elucidated the driving mechanism of reconstruction by combining density functional theory (DFT) with high-precision machine-learning interatomic potentials. By performing high-throughput energy mapping across the continuous orientation space, we revealed that the experimentally observed high-index facets—such as {113}, {610}, and {711}—precisely correspond to local minima in the surface energy landscape. This independent agreement between experimental observation and theoretical prediction constitutes a robust double-blind test, confirming that the reconstruction is fundamentally an energy-driven, orientation-selective process.

Ultimately, this study provides the first systematic resolution of all possible reconstructed facets for the graphene-copper system and establishes a unified relationship between orientation selectivity and thermodynamic stability. Beyond this specific system, the proposed framework—using non-Euclidean topologies to probe continuous parameter spaces—offers a generalizable paradigm for exploring and controlling interfacial phenomena in diverse metal–2D material heterostructures.

## Method

### *In Situ* CVD Growth

*In situ* CVD growth experiments were performed inside the chamber of a modified commercial ESEM (Quattro ESEM, Thermo Fisher). The vacuum system of the ESEM was modified and upgraded with oil-free pre-vacuum pumps. The instrument is equipped with a home-made laser heating stage, a gas supply unit (using mass flow controllers from Bronkhorst) and a mass spectrometer (OmniStar, Pfeiffer) for analysis of the chamber atmosphere. After each sample loading, the chamber was pumped out to around $10^{-3}$ Pa, purged with pure $N_2$, and pumped again to $10^{-3}$ Pa successively several times. Under the CVD growth conditions, the pressure is six orders of magnitude higher than the base pressure and is made up mostly of $H_2$ (99.9995% purity) and $C_2H_4$ (99.95% purity). Polycrystalline copper foils from Alfa Aesar were used as substrate (99.999% purity). Prior to all CVD growth experiments, the chamber of the ESEM was plasma cleaned. All samples were annealed at 1000 °C under a hydrogen flow of 8 sccm at a pressure of around $4.4 \times 10^{-2}$ Pa for 50 min inside the chamber. The temperature was measured using a type B thermocouple that was spot-welded onto the substrate, which simultaneously served to ground the sample. CVD growth was performed at 1000 °C using a flow of 4 sccm $H_2$ and 0.1 sccm of $C_2H_4$ at a total chamber pressure of $(2–4) \times 10^{-2}$ Pa. During the experiments, the microscope was operated at an acceleration voltage of 5.0 kV. Images were recorded using the secondary electron signal collected by a standard Everhart Thornley detector (ETD) during sample annealing, CVD growth, and cooling. Images were recorded using a large-field detector during the CVD growth conditions. No influence of the electron beam on the growth and etching process could be observed. The imaged regions and their respective surroundings showed similar behavior, as demonstrated by changing the magnification or by moving the sample under the beam. Furthermore, no electron-beam-induced contamination was observed at elevated temperatures.

**Electron backscatter diffraction**

Electron backscatter diffraction (EBSD) patterns were acquired using the EDAX Velocity Plus detector installed on a Thermo Fisher Apreo microscope. Analysis of EBSD patterns, including phase identification and generation of orientation maps, was performed with the EDAX OMI 8.6.108 program.

**Atomic force microscopy**

Images were taken on a Bruker Dimension Fastscan. Imaging was done in tapping mode using a V-shaped cantilever Tap300Al-G (silicon-tip on Nitride Lever with frequency $f_0$ = 200–400 kHz and spring constant $k$ = 20-75 N/m).

**Cross-sectional sample preparation**

The Gra-Cu cross section sample was prepared by focus ion beam technology in Thermal Scientific Helios 5 CX DualBeam. To protect the surface of interest, a composite carbon layer was deposited, consisting of an initial electron-beam-induced deposition (2 kV, 3.2 nA, ~200 nm) followed by ion-beam-induced deposition (30 kV, 0.5 nA). Coarse trenching was performed using ion beam currents decreasing from 15 nA to 3 nA. Subsequently, the lamella was gradually thinned to a final thickness of less than 100 nm by reducing the ion beam current from 1 nA to 50 pA. Finally, to remove the amorphous damage layer, a low-voltage polishing step was conducted at 5 kV with the beam current reducing from <100 pA to ~10 pA.

**Transmission electron microscopy**

High-resolution transmission electron microscopy and scanning transmission electron microscopy were carried out using a JEOL Grand Arm 300F with double Aberration-correctors. All the TEM/STEM images are taken at 300kV in order to get clear atomic structure of Cu substrate and the steps in the Gra-Cu interface. OneView camera binned to 2k × 2k pixel resolution was used to acquire images, including the real space and selective electron diffraction patterns in TEM mode and the Kikuchi patterns in Ronchigram. All transmission electron microscope images were then post-processed through Gatan Microscopy Suite software.

1   Wang, L. *et al.* Epitaxial growth of a 100-square-centimetre single-crystal hexagonal boron nitride monolayer on copper. *Nature* **570**, 91-95 (2019). https://doi.org:10.1038/s41586-019-1226-z

2   Li, X. *et al.* Large-Area Synthesis of High-Quality and Uniform Graphene Films on Copper Foils. *Science* **324**, 1312-1314 (2009). https://doi.org:10.1126/science.1171245

3   Luo, R. *et al.* Van der Waals interfacial reconstruction in monolayer transition-metal dichalcogenides and gold heterojunctions. *Nature Communications* **11**, 1011 (2020). https://doi.org:10.1038/s41467-020-14753-8

4   Jiang, H. *et al.* Two-dimensional Czochralski growth of single-crystal MoS2. *Nature Materials* **24**, 188-196 (2025). https://doi.org:10.1038/s41563-024-02069-7

5   Deng, B. *et al.* Wrinkle-Free Single-Crystal Graphene Wafer Grown on Strain-Engineered Substrates. *ACS Nano* **11**, 12337-12345 (2017). https://doi.org:10.1021/acsnano.7b06196

6   Gao, L. *et al.* Repeated growth and bubbling transfer of graphene with millimetre-size single-crystal grains using platinum. *Nature Communications* **3**, 699 (2012). https://doi.org:10.1038/ncomms1702

7   Kim, J. Y. *et al.* High Facets on Nanowrinkled Cu via Chemical Vapor Deposition Graphene Growth for Efficient CO2 Reduction into Ethanol. *ACS Catalysis* **11**, 5658-5665 (2021). https://doi.org:10.1021/acscatal.0c05263

8   Emtsev, K. V. *et al.* Towards wafer-size graphene layers by atmospheric pressure graphitization of silicon carbide. *Nature Materials* **8**, 203-207 (2009). https://doi.org:10.1038/nmat2382

9   Ni, G.-X. *et al.* Quasi-Periodic Nanoripples in Graphene Grown by Chemical Vapor Deposition and Its Impact on Charge Transport. *ACS Nano* **6**, 1158-1164 (2012). https://doi.org:10.1021/nn203775x

10  Geng, D. *et al.* Large-Area Growth of Five-Lobed and Triangular Graphene Grains on Textured Cu Substrate. *Advanced Materials Interfaces* **3**, 1600347 (2016). https://doi.org:https://doi.org/10.1002/admi.201600347

11  Tian, J. *et al.* Graphene Induced Surface Reconstruction of Cu. *Nano Letters* **12**, 3893-

| | |
|---|---|
| | 3899 (2012). https://doi.org:10.1021/nl3002974 |
| 12 | Hedayat, S. M., Karimi-Sabet, J. & Shariaty-Niassar, M. Evolution effects of the copper surface morphology on the nucleation density and growth of graphene domains at different growth pressures. *Applied Surface Science* **399**, 542-550 (2017). https://doi.org:https://doi.org/10.1016/j.apsusc.2016.12.126 |
| 13 | Wang, Z.-J. *et al.* Direct Observation of Graphene Growth and Associated Copper Substrate Dynamics by in Situ Scanning Electron Microscopy. *ACS Nano* **9**, 1506-1519 (2015). https://doi.org:10.1021/nn5059826 |
| 14 | Kim, D. W., Lee, J., Kim, S. J., Jeon, S. & Jung, H.-T. The effects of the crystalline orientation of Cu domains on the formation of nanoripple arrays in CVD-grown graphene on Cu. *Journal of Materials Chemistry C* **1**, 7819-7824 (2013). https://doi.org:10.1039/C3TC31717J |
| 15 | Kang, J. H. *et al.* Strain Relaxation of Graphene Layers by Cu Surface Roughening. *Nano Letters* **16**, 5993-5998 (2016). https://doi.org:10.1021/acs.nanolett.6b01578 |
| 16 | Deng, B. *et al.* Anisotropic Strain Relaxation of Graphene by Corrugation on Copper Crystal Surfaces. |
| 17 | Yi, D. *et al.* What Drives Metal-Surface Step Bunching in Graphene Chemical Vapor Deposition? *Physical Review Letters* **120**, 246101 (2018). https://doi.org:10.1103/PhysRevLett.120.246101 |
| 18 | Šrut Rakić, I., Čapeta, D., Plodinec, M. & Kralj, M. Large-scale transfer and characterization of macroscopic periodically nano-rippled graphene. *Carbon* **96**, 243-249 (2016). https://doi.org:https://doi.org/10.1016/j.carbon.2015.09.046 |
| 19 | Zhang, H. *et al.* Stripe distributions of graphene-coated Cu foils and their effects on the reduction of graphene wrinkles. *RSC Advances* **5**, 96587-96592 (2015). https://doi.org:10.1039/C5RA17581J |
| 20 | Schädlich, P. *et al.* Stacking Relations and Substrate Interaction of Graphene on Copper Foil. *Advanced Materials Interfaces* **8** (2021). https://doi.org:10.1002/admi.202002025 |
| 21 | Nie, S. *et al.* Growth from below: bilayer graphene on copper by chemical vapor deposition. *New Journal of Physics* **14**, 093028 (2012). https://doi.org:10.1088/1367-2630/14/9/093028 |
| 22 | Vondráček, M. *et al.* Nanofaceting as a stamp for periodic graphene charge carrier modulations. *Scientific Reports* **6**, 23663 (2016). https://doi.org:10.1038/srep23663 |
| 23 | Rasool, H. I. *et al.* Continuity of Graphene on Polycrystalline Copper. *Nano Letters* **11**, 251-256 (2011). https://doi.org:10.1021/nl1036403 |
| 24 | Hayashi, K., Sato, S. & Yokoyama, N. Anisotropic graphene growth accompanied by step bunching on a dynamic copper surface. *Nanotechnology* **24**, 025603 (2013). https://doi.org:10.1088/0957-4484/24/2/025603 |
| 25 | Deng, B. *et al.* Growth of Ultraflat Graphene with Greatly Enhanced Mechanical Properties. *Nano Letters* **20**, 6798-6806 (2020). https://doi.org:10.1021/acs.nanolett.0c02785 |
| 26 | Wang, Y. *et al.* Ultraflat single-crystal hexagonal boron nitride for wafer-scale integration of a 2D-compatible high-κ metal gate. *Nature Materials* **23**, 1495-1501 (2024). https://doi.org:10.1038/s41563-024-01968-z |
| 27 | Lui, C. H., Liu, L., Mak, K. F., Flynn, G. W. & Heinz, T. F. Ultraflat graphene. *Nature* **462**, 339-341 (2009). https://doi.org:10.1038/nature08569 |

28  Zhao, Y. *et al.* Supertwisted spirals of layered materials enabled by growth on non-Euclidean surfaces. *Science* **370**, 442-445 (2020). https://doi.org:10.1126/science.abc4284

29  Kraus, J. *et al.* Towards the perfect graphene membrane? – Improvement and limits during formation of high quality graphene grown on Cu-foils. *Carbon* **64**, 377-390 (2013). https://doi.org:https://doi.org/10.1016/j.carbon.2013.07.090

30  Surana, M. *et al.* Strain-Driven Faceting of Graphene-Catalyst Interfaces. *Nano Letters* **23**, 1659-1665 (2023). https://doi.org:10.1021/acs.nanolett.2c03911

31  Ananthakrishnan, G. *et al.* Graphene-mediated stabilization of surface facets on metal substrates. *Journal of Applied Physics* **130**, 165302 (2021). https://doi.org:10.1063/5.0065107

32  Chen, S. *et al.* Millimeter-Size Single-Crystal Graphene by Suppressing Evaporative Loss of Cu During Low Pressure Chemical Vapor Deposition. *Advanced Materials* **25**, 2062-2065 (2013). https://doi.org:https://doi.org/10.1002/adma.201204000

33  Li, X. *et al.* Large-Area Graphene Single Crystals Grown by Low-Pressure Chemical Vapor Deposition of Methane on Copper. *Journal of the American Chemical Society* **133**, 2816-2819 (2011). https://doi.org:10.1021/ja109793s

34  Vlassiouk, I. *et al.* Graphene Nucleation Density on Copper: Fundamental Role of Background Pressure. *The Journal of Physical Chemistry C* **117**, 18919-18926 (2013). https://doi.org:10.1021/jp4047648

35  Vlassiouk, I. *et al.* Large scale atmospheric pressure chemical vapor deposition of graphene. *Carbon* **54**, 58-67 (2013). https://doi.org:https://doi.org/10.1016/j.carbon.2012.11.003

36  Wu, T. *et al.* Fast growth of inch-sized single-crystalline graphene from a controlled single nucleus on Cu–Ni alloys. *Nature Materials* **15**, 43-47 (2016). https://doi.org:10.1038/nmat4477

37  Xu, X. *et al.* Ultrafast growth of single-crystal graphene assisted by a continuous oxygen supply. *Nature Nanotechnology* **11**, 930-935 (2016). https://doi.org:10.1038/nnano.2016.132

38  Loginova, E., Bartelt, N. C., Feibelman, P. J. & McCarty, K. F. Evidence for graphene growth by C cluster attachment. *New Journal of Physics* **10**, 093026 (2008). https://doi.org:10.1088/1367-2630/10/9/093026

39  Chen, S. *et al.* Oxidation Resistance of Graphene-Coated Cu and Cu/Ni Alloy. *ACS Nano* **5**, 1321-1327 (2011). https://doi.org:10.1021/nn103028d

40  Yuan, G. *et al.* Proton-assisted growth of ultra-flat graphene films. *Nature* **577**, 204-208 (2020). https://doi.org:10.1038/s41586-019-1870-3

41  Hazen, R. M. & Sholl, D. S. Chiral selection on inorganic crystalline surfaces. *Nature Materials* **2**, 367-374 (2003). https://doi.org:10.1038/nmat879

42  Vitos, L., Ruban, A. V., Skriver, H. L. & Kollár, J. The surface energy of metals. *Surface Science* **411**, 186-202 (1998). https://doi.org:https://doi.org/10.1016/S0039-6028(98)00363-X

43  Agrawal, A., Raskar, R. & Chellappa, R. in *Computer Vision – ECCV 2006.* (eds Aleš Leonardis, Horst Bischof, & Axel Pinz) 578-591 (Springer Berlin Heidelberg).

44  Simchony, T., Chellappa, R. & Shao, M. Direct analytical methods for solving Poisson equations in computer vision problems. *IEEE Transactions on Pattern Analysis and*

|     | *Machine Intelligence* **12**, 435-446 (1990). https://doi.org:10.1109/34.55103 |
| --- | --- |
| 45  | Harker, M. & Leary, P. O. in *2008 IEEE Conference on Computer Vision and Pattern Recognition.*  1-7. |
| 46  | Kang, M. & SUN, G.-M. |
| 47  | Bae, J., Chatzidakis, S. & Bean, R. in *International Conference on Nuclear Engineering.*  V004T014A011 (American Society of Mechanical Engineers). |
| 48  | Wulff, G. XXV. Zur Frage der Geschwindigkeit des Wachsthums und der Auflösung der Krystallflächen.  **34**, 449-530 (1901). https://doi.org:doi:10.1524/zkri.1901.34.1.449 |
| 49  | Herring, C. Some Theorems on the Free Energies of Crystal Surfaces. *Physical Review* **82**, 87-93 (1951). https://doi.org:10.1103/PhysRev.82.87 |
| 50  | Chen, C. *et al.* Large local lattice expansion in graphene adlayers grown on copper. *Nature Materials* **17**, 450-455 (2018). https://doi.org:10.1038/s41563-018-0053-1 |
| 51  | Niehus, H. Surface reconstruction of Cu (111) upon oxygen adsorption. *Surface Science* **130**, 41-49 (1983). https://doi.org:https://doi.org/10.1016/0039-6028(83)90258-3 |
| 52  | Wuttig, M., Franchy, R. & Ibach, H. Oxygen on Cu(100) – a case of an adsorbate induced reconstruction. *Surface Science* **213**, 103-136 (1989). https://doi.org:https://doi.org/10.1016/0039-6028(89)90254-9 |
| 53  | Jensen, F., Besenbacher, F. & Stensgaard, I. Two new oxygen induced reconstructions on Cu(111). *Surface Science* **269-270**, 400-404 (1992). https://doi.org:https://doi.org/10.1016/0039-6028(92)91282-G |
| 54  | Witte, G. *et al.* Oxygen-induced reconstructions on Cu(211). *Physical Review B* **58**, 13224-13232 (1998). https://doi.org:10.1103/PhysRevB.58.13224 |
| 55  | Jensen, F., Besenbacher, F., Laegsgaard, E. & Stensgaard, I. Dynamics of oxygen-induced reconstruction of Cu(100) studied by scanning tunneling microscopy. *Physical Review B* **42**, 9206-9209 (1990). https://doi.org:10.1103/PhysRevB.42.9206 |
| 56  | Wang, H. *et al.* Controllable Synthesis of Submillimeter Single-Crystal Monolayer Graphene Domains on Copper Foils by Suppressing Nucleation. *Journal of the American Chemical Society* **134**, 3627-3630 (2012). https://doi.org:10.1021/ja2105976 |